\documentclass[a4paper, fleqn, usenatbib, useAMS]{tex/mnras}

\usepackage{newtxtext, newtxmath}
\usepackage[T1]{fontenc}
\usepackage{ae, aecompl}

\usepackage{comment}

\usepackage{graphicx}
\graphicspath{{figs/}}
\DeclareGraphicsExtensions{.pdf, .ps, .eps}
\usepackage[multidot]{grffile}

\usepackage{xspace}
\usepackage[usenames, dvipsnames]{xcolor}

\newcommand{\starlight}{\textsc{starlight}\xspace}

\title[Rebutting fake news on full spectral fitting] 
      {Rebutting fake news on full spectral fitting}

\author[R. Cid Fernandes]
       {Roberto Cid Fernandes$^{1}$\thanks{e-mail:cid@astro.ufsc.br}
        \\
        $^{1}$Departamento de F\'{\i}sica--CFM, Universidade Federal de Santa Catarina, C.P.\ 476, 88040-900, Florian\'opolis, SC, Brazil \\
       }

\date{Accepted July 25. Received June 1; in original form June 1}
\pubyear{2018}

\begin{document}

\label{firstpage}
\pagerange{\pageref{firstpage}--\pageref{lastpage}}

\maketitle

%**********************************************************************%
%                                                                      %
%                               Abstract                               %
%                                                                      %
%**********************************************************************%
\begin{abstract}
A recent paper by Ge et al.\ performs a series of experiments with two full spectral fitting codes, pPXF and \starlight, finding that the two yield consistent results when the input spectrum is not heavily reddened. For $E(B-V) \ga 0.2$, however, they claim \starlight\ leads to severe biases in the derived properties. Counterintuitively, and at odds with previous simulations, they find that this behaviour worsens significantly as the signal-to-noise ratio of the input spectrum increases.
This  communication shows that this is entirely due to an $A_V < 1$ mag condition imposed while initializing the Markov chains in the code. This choice is normally irrelevant in real-life galaxy work but can become critical in artificial experiments. Alleviating this usually harmless initialization constraint changes the Ge et al.\ results completely, as was explained to the authors before their publication.
We replicate their spectral fitting experiments, finding much smaller biases. Furthermore both bias and scatter in the derived properties all converge as $S/N$ increases, as one would expect. We also show how the very output of the code provides ways of diagnosing anomalies in the fits. The code behaviour has been documented in careful and extensive experiments in the literature, but the biased analysis of Ge et al.\ is just not representative of \starlight\ at all.
\end{abstract}


%%%%%%%%%%%%%%%%%%%%%%%%%%%%%%%%%%%%%%%%%%%%%%%%%%%%%%%%%%%%%%%%%%%%%%%%%%%%%%%%%%%%%%%%%%%%%%%%%%%%%%%%%%%%%%

\section{Introduction}
\label{sec:intro}

Full spectral synthesis techniques have blossomed after \cite{Bruzual2003} released a suite of evolutionary population synthesis models for the spectra of simple stellar populations (SSP) as a function of age and metallicity. That paper updated the theory to a spectral resolution which  observational work on galaxies had achieved long before, fostering the development of methods to match observed galaxy spectra with combinations of the SSP models. These methods became generally known as ``full spectral fitting'', highlighting the $\lambda$-by-$\lambda$ nature of how data and models are compared.

Reviews on this subject can be found in \cite{Walcher2011} and \cite{Conroy2013}, while \cite{CidFernandes2006BAAA,CidFernandes2007IAU} gives outdated but useful reviews reflecting the early days of the field. The introduction section in the recent paper by \citet[hereafter GYCMLL]{Ge2018} is also a good source of references. The authors then proceed to compare two publicly available full spectral fitting codes: pPXF \citep{Cappellari2004,Cappellari2017}, and {\sc starlight} \citep{CidFernandes2005}. 

Comparing different methods is a tedious yet useful exercise if carried out with due care. Unfortunately, GYCMLL completely misrepresent the performance of {\sc starlight}. Their results convey the false idea that the code leads to significant biases in derived properties such as $E(B-V)$, mean stellar age, and metallicity. Particularly disastrous results are obtained for young and dusty systems, specially when assigned exquisite signal-to-noise ratios.

The sole purpose of this communication is to set the record straight. We present a revised version of the same spectral fitting experiments performed by GYCMLL. The results, however, could hardly be more different. The nature of the difference lies on an irrelevant and trivially fixable technicality which GYCMLL were fully aware of. 
As a result they end up painting a distorted and biased picture of \starlight's performance. This calls for clarification, and this is what is presented here.

\section{Spectral fitting experiments}
\label{sec:analysis}

The experiments performed by GYCMLL follow the usual Monte Carlo logic of: (1) take an input spectrum of known properties, (2) perturb it with  gaussian noise, (3) process the perturbed spectrum through the spectral fitting code, and (4) compare the output properties  with the input ones.

Their main set of experiments uses solar metallicity  SSP spectra with five different ages ($\log t/{\rm yr} = 8.0$, 8.5, 9.0, 9.5, 10.0) and attenuated by a foreground screen of dust producing $E(B-V)$ values of 0.0, 0.1, 0.2, 0.3, 0.4 and 0.5 (the \citealt{Calzetti2000} reddening curve is adopted). The stellar population library is that of \cite{Vazdekis2010} for a Salpeter IMF and ``Padova 2000" evolutionary tracks \citep{Girardi2000}.  Each of these 30 input spectra (5 ages $\times$ 6 extinctions) was then perturbed with gaussian noise with a $\lambda$-independent amplitude  adjusted to produce signal-to-noise ratios of $S/N = 10$, 18, 32, 56, 100, 178, and 316 in the 5490--5510 \AA\ interval. 
In order to estimate the biases and uncertainties in derived properties 50 incarnations of each input spectrum were built, each with a different realization of the noise. 

We have replicated these simulations, generating our own set of mock spectra exactly as described above. We then fit them with \starlight\  using a spectral base of 150 SSPs spanning 25 logarithmically-spaced ages between $t = 63$ Myr and 15.8 Gyr and 6 metallicities ($[M/H] = -1.7$, -1.3, -0.7, -0.4, 0.0, and $+0.2$) also from 
\cite{Vazdekis2010}, and assuming a Calzetti extinction curve, as done by GYCMLL.

\subsection{Results}
\label{sec:Results}

%------------------------------- Figure -------------------------------%
\begin{figure*}
  \includegraphics[width=\textwidth, clip=true, trim = 0 170 0 0]{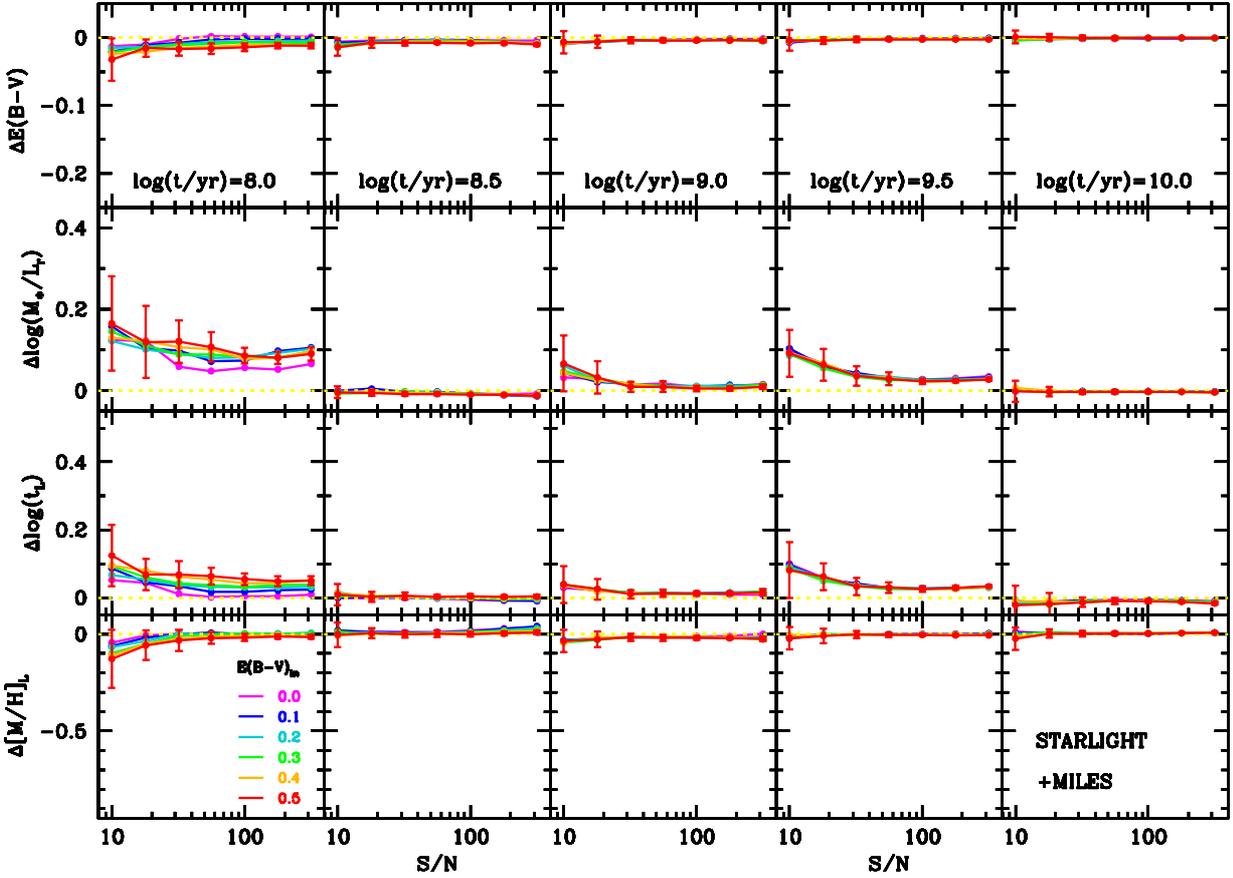}
  \caption{A revised version of figure 2 in GYCMLL. The different rows show the output minus input biases in $E(B-V)$, mass-to-light ratio, mean stellar age, and metalicity as a function of $S/N$. Each column shows results for a different input age, indicated in the top row. Colours code the input value of $E(B-V)$, as labeled in the bottom-left panel. For clarity, $\pm 1 \sigma$ error bars are only shown for the extreme case of $E(B-V)_{\rm in}=0.5$  (red). 
}
\label{fig:GeFig2}
\end{figure*}
%------------------------------- Figure -------------------------------%

Fig.\ \ref{fig:GeFig2} shows our results.  The plot has the same layout and axis scales as figure 2 of GYCMLL, to which it should be compared. Results for the five input ages are shown in different columns (as labeled in the top row), while the input $E(B-V)$ values are coded by different colors, as indicated in the bottom left panel. 

Panels in the top row show the difference between the output and input values of $E(B-V)$. The curves connect the mean values of the bias $\Delta E(B-V) = E(B-V)_{\rm out} - E(B-V)_{\rm in}$ obtained from the 50 perturbed versions of each input spectrum. Error bars are only shown for the runs with $E(B-V)_{\rm in} = 0.5$ (in red), but the scatter is about the same for other values of $E(B-V)_{\rm in}$.

The contrast with the results of GYCMLL is striking. Take the red curve for the 100 Myr model, for instance. While they obtain a bias of $\Delta E(B-V) = -0.20 \pm 0.02$ at $S/N = 316$ we obtain an insignificant $\Delta E(B-V) = 
-0.011 \pm  0.005$. Moreover, as expected on the basis of pure common sense, but contrary to what they find, the bias decreases as the $S/N$ improves.

Other panels in Fig.\ \ref{fig:GeFig2} show the statistics of the output minus input values of the mass to light ratio, the luminosity weighted mean age and metallicity. The results again thoroughly contradict those reported by GYCMLL. Focusing again on the runs for $t_{\rm in} = 100$ Myr and $E(B-V)_{\rm in} = 0.5$,  they find that $\Delta \log(t_L)$ (see equation 3 of their paper)  starts from $\sim 0.3$ dex at $S/N = 10$ and grows to $\sim 0.5$ dex at $S/N = 316$. Similarly, the bias in metallicity starts off badly and only worsens as the $S/N$ improves. The real situation is shown by the red curves in the first column of Fig.\ \ref{fig:GeFig2}, which show modest biases of about 0.1 dex at low $S/N$ that decrease to negligible values as the $S/N$ increases.

%------------------------------- Figure -------------------------------%
\begin{figure*}
  \includegraphics[width=\textwidth, clip=true, trim = 0 85 0 0]{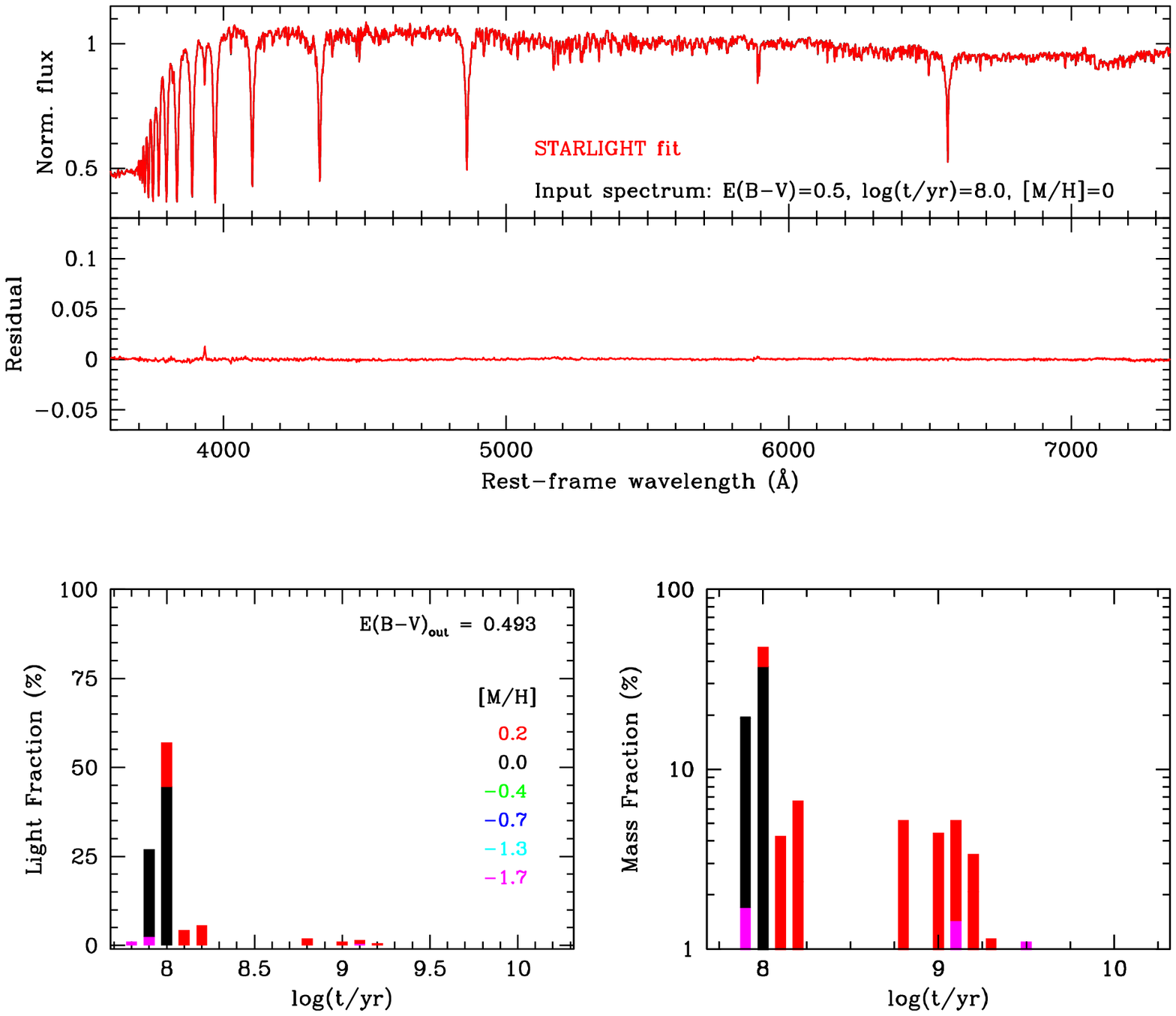}
  \caption{A revised version of figure 5 in GYCMLL. The top panels show the input spectrum (in black) of a solar metalicity 100 Myr old SSP extincted by $E(B-V)=0.5$,  the $\sim$ indistinguishable  \starlight\ fit (in red), and the corresponding output minus input residual spectrum on a zoomed scale. 
Bottom panels shows the mixture of SSP spectra behind the model fit in terms of light (left) and mass (right) fractions.}
\label{fig:GeFig5K0}
\end{figure*}
%------------------------------- Figure -------------------------------%

Before explaining the reason for these dramatic differences let us also replicate their figure 5, where they show that \starlight\ fails to fit the spectrum and recover the parameters of an input 100 Myr, solar metallicity SSP reddened by $E(B-V)_{\rm in} = 0.5$. Fig.\ \ref{fig:GeFig5K0} shows our version of that plot.
The top panel shows the input spectrum (in black) and the \starlight\ fit in red. The difference between the two is only visible in the residual spectrum (shown in the panel below), and even then it is hardly noticeable. In the units of the plot, where the flux at 5635 \AA\ is $\equiv 1$,  the rms of the residual is  a tiny 
0.0009, and even the peak value is just 
0.0128 (over the CaII K line). The fit shown by GYCMLL is much worse, with residuals an order of magnitude larger.

The bottom panels in Fig.\ \ref{fig:GeFig5K0} show the corresponding \starlight\ solution in terms of light and mass fractions. The light distribution (at the 5635 \AA\ normalization wavelength, left panel) peaks at the correct age, and 88\% of it is contained within $\pm 0.1$ dex of $\log(t/{\rm yr}) = 8.0$. The output  $E(B-V)$ is  0.493, and the metallicity is mostly solar (73\% of the light in $[M/H] = 0.0$ and most of the rest in $[M/H] = +0.2$). Not identical to the input parameters, but pretty close by any real-world standard. This fit is obtained with the same configuration file used by GYCMLL\footnote{GYCMLL incorrectly associate this configuration file with that used in the  ``state-of-the-art analysis of the CALIFA dataset by \cite{Amorim2017}''. That would not even be possible, as the version of \starlight\ used in that paper differs from the public one in several details, including the configuration file.}. There are several ways to play with the configuration parameters to further improve the fit (see Fig. \ref{fig:GeFig5K2}), although one should wonder how necessary or relevant it is to better a fit which already produces such tiny residuals. Also, as discussed in Section \ref{sec:SCF}, the code itself points to a better solution.

The bottom right panel translates light to mass fractions. Again, the distribution peaks in the input population. Inevitably, the scatter is larger, since the input model is nearly the youngest in the base, with a $M/L_{\lambda5635}$ ratio dozens of times smaller than the oldest ones. This explains why the insignificant 1\% of light attributed to the 1 Gyr, $[M/H] = +0.2$ population gets inflated to $5\%$ of the mass, and similarly for the other components. The same effect results in a bias of about $+0.1$ dex in the mass-to-light ratio even at high $S/N$, as seen in the first column of  the $\Delta \log M/L_r$ panels in Fig.\ \ref{fig:GeFig2}. As should be evident from its very name, \starlight\ fits light, not mass fractions. It should be equally evident even to non \starlight\ users that components accounting for such tiny light fractions should be treated as noise, not signal.\footnote{The analysis of which components are significant is usually done a posteriori, while analyzing the output of a fit, but the code also offers ways of setting a priori significance thresholds by adjusting the corresponding parameters in the  configuration file.}

The comparison of our Fig.\ \ref{fig:GeFig5K0} with figure 5 in GYCMLL reveals drastic differences.  
In their version the \starlight\ solution is scattered all over the age-metallicity plane. They only show results in terms of mass fractions, which boosts the scatter due to the highly non-linear $M$-$L$  relation of stars. In terms of light fractions their results would not look as scattered, though still much worse than those seen in the bottom left panel of  Fig.\ \ref{fig:GeFig5K0}.

How come the results reported here are so strikingly different from those of GYCMLL?

\subsection{The futile reason for the discrepancy}
\label{sec:TheReason}

The reason for these completely different results lies on a hitherto deemed unimportant line of the code which stipulates a maximum $A_V$ of 1 mag while initializing the Markov chains, instead of allowing it to reach the upper limit established in the configuration file.
The logic behind this single if/then line of the code  was that real galaxies seldom suffer so much extinction, so starting the parameter chains from $A_V$ as large as 4  (as in the example configuration file distributed with the code and used by GYCMLL) would be a waste of time and an unnecessary hindrance to the progress of the chains.

In any case, if a spectrum does require more dust to be fitted the chains should travel to high $A_V$ regions of the parameter space even if initialized at low values. This indeed happens in the $E(B-V)_{\rm in} \ge 0.3$ (equivalent to $A_{V,{\rm in}} \ge 1.2$) \starlight\ fits of GYCMLL, which do cross the $A_V < 1$ limit of the initial chains. In some of their simulations, however, the chains do not run for long enough to reach the correct values.
Since $A_V / E(B-V) = 4.05$ for the reddening law used in these experiments, $A_V < 1$ implies $E(B-V) < 0.25$, so that their conclusion that \starlight\ works well when $E(B-V) \la 0.2$ is a mere reflection of this initialization strategy.

The difference between our figures and their versions in GYCMLL boils down to removing this single line of the code. Clearly this otherwise harmless and innocent technical condition has nefarious effects for this particular experiment. 

The reason why their results are particularly disastrous when young and highly reddened SSPs are used as input to the code is rather technical, though not difficult to grasp. Old SSPs are already naturally red, and can only get substantially redder by increasing $A_V$, so the code recovers their properties even if the chains start far from the right parameters. A highly reddened young population, however, has a spectrum whose overall shape can be well matched by combinations of older stellar populations with less $A_V$. Indeed, even the bad fit example shown by GYCMLL does match the continuum shape, failing only in the absorption lines.
These regions of the parameter space are visited first by the Markov chains, specially if they start from low $A_V$. In order to reach the extreme corner of the parameter space containing the true solution, the chains must be allowed to run until they realize there are solutions where not only the continuum shape can be fitted, but also the absorption lines. 
The setup parameters in the example configuration files provided in the public distribution of \starlight\ do not let the chains run for long enough to reach $A_V$ values as high as needed in this case.
The higher the $S/N$ the longer it takes to reach this region, which explains why GYCMLL find the counterintuitive and ilogical result that biases worsen as $S/N$ increases, as well as the large computing times.\footnote{Inspection of the (usually ignored) run-time screen output of the code reveals signs of anomalous behaviour of the chains, like very small step sizes, and in some cases warnings that the ``rapid $\chi^2$ tricks'' explained in the user manual are struggling.}

Though it is possible to overcome the biases reported by GYCMLL by playing with the several technical knobs and weighting options provided by \starlight\ (an alternative explored in their so called ``slow mode'' fits, also explored here in appendix \ref{app:AppendixA}), simply removing the $A_V < 1$ condition already leads to perfectly satisfactory results. Indeed, Figs.\ \ref{fig:GeFig2} and  \ref{fig:GeFig5K0} use the same configuration used by GYCMLL.

All of this was made abundantly clear to Ge at al.\ through private communications with the third author in December/2017. Their paper does in fact echo some of the explanations given then and expanded here. Instead of using the revised code, however, they chose to focus their whole study on experiments which exploit what is in practice an irrelevant and trivially fixable technical vulnerability of the  code. The result is a biased mischaracterization of \starlight's performance which serves no purpose but to cloud the field with meaningless simulations.\footnote{The running terminology for this is ``fake news''. }
 
The revised version of the code is available at www.starlight.ufsc.br for users who suspect their analysis may have been affected by the same issue. As discussed in section \ref{sec:TestYourRun}, where we test to which extent previous fits of SDSS and CALIFA spectra were affected by this, the likelihood of this happening in real galaxy work is tiny, so most users should not worry about this issue. If the object under study is as extreme as the worst cases explored by GYCMLL, where all the light is concentrated in a highly reddened single population, then biases like those they report may indeed happen. However, if this is your case, then you should be using \starlight\ in a different way. In fact, as explained next, the output of the original code already contains a much better solution than the one used by GYCMLL.

\subsection{Single component fits}
\label{sec:SCF}

%------------------------------- Figure -------------------------------%
\begin{figure}
  \includegraphics[width=0.5\textwidth, clip=true, trim = 0 20 0 0]{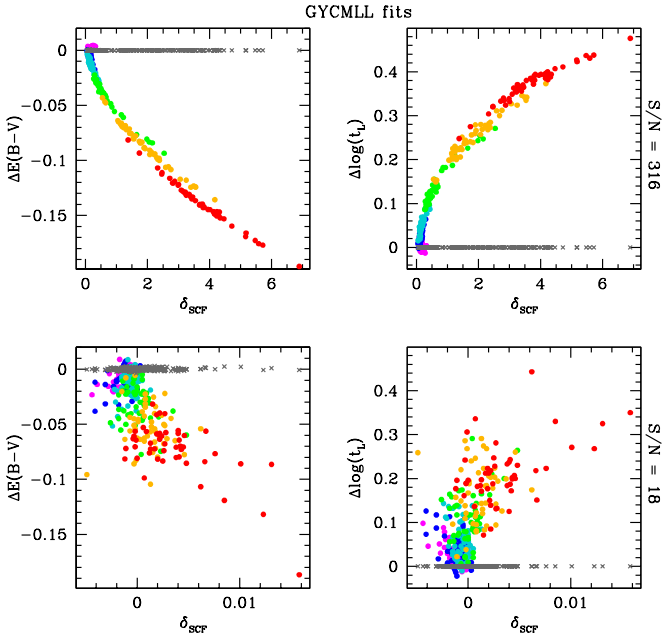}
  \caption{Biases in color excess and age (as derived from simulations like those of GYCMLL and with the unrevised code) against the ``problem detector" $\delta_{\rm SCF}$ index. Positive values of $\delta_{\rm SCF}$ indicate that a single component provides a better spectral fit than the one found for a mixed stellar population. All points are for an input spectrum composed of an SSP of 100 Myr. The $S/N$ is 316 and 18 in the top and bottom panels, respectively. Circles are coloured according to $E(B-V)_{\rm in}$ following the same pallete shown in the bottom left panel of Fig.\ \ref{fig:GeFig2}. Gray crosses mark the best single component fit. Mixed population solutions producing $\delta_{\rm SCF} > 0$ should be rejected.}
\label{fig:chi2SSPTest}
\end{figure}
%------------------------------- Figure -------------------------------%

As documented in its user manual, besides the light fractions associated with the smallest $\chi^2$ model found by the Markov chains (the ``official'' best solution), \starlight\ also fits the input spectrum with each of the base components individually. In principle this feature should only interest users working with star clusters, for whom the question is not what are the mixture parameters, but what is the single component which best fits the data and what is the implied extinction. These single component fits (SCF) can however be used to perform a very basic test of the reasonability of the best fit mixed model. 

The test goes like this: If any of the $\chi^2_{\rm SCF}$'s is smaller than the $\chi^2$ of the best-fit mixed population, then either you are fitting a star cluster or something is not right (e.g., the chains have not run for long enough, problems with  the input data, etc.). In either case, when $\chi^2_{\rm SCF} < \chi^2$ the mixed model should either be ruled out or replaced by the best SCF.

Had GYCMLL applied this sanity check to their simulations they would have rejected most of their \starlight\ fits, particularly those at high $S/N$.  To demonstrate this we have run the same simulations as before, but now with the original code with the problematic $A_V < 1$ initialization.

The results are shown  in Fig.\ \ref{fig:chi2SSPTest}, where we plot the bias in $E(B-V)$ and $\log(t_L)$ against $\delta_{\rm SCF} \equiv (\chi^2 - \chi^2_{\rm SCF}) / N$, where $N$ is the number of wavelengths. This index measures how much worse the mixed solution is with respect to the SCF. For clarity only results for $t_{\rm in} = 100$ Myr (the worst case) are shown. Top panels are for runs with $S/N = 316$ (the largest in their simulations)  and bottom ones for $S/N = 18$. Circles are coloured according to the input value of $E(B-V)$ as in Fig.\ \ref{fig:GeFig2}. As reported by CYGMLL, biases are larger the larger $E(B-V)_{\rm in}$ is. Gray crosses show the difference between the best SCF and input properties. As expected, the SCF recover the input parameters exactly.\footnote{Technically speaking, \starlight\ did give GYCMLL the correct solution. It just so happens that because of the particularities of their simulations this solution was not in the best mixed model column of the output, but in  the one reporting the single component fits.}

At high $S/N$ all points have $\delta_{\rm SCF} > 0$, signaling that  the single component, and not the mixed population fit should be considered the best solution. More importantly, the correspondence between $\delta_{\rm SCF}$ and the bias is evident, confirming that this index can be used to track problematic fits.
At lower $S/N$ $\delta_{\rm SCF}$ also tracks the bias, but with more scatter. The values of $\delta_{\rm SCF}$ are also much smaller because the $\chi^2$'s scale as $(S/N)^2$. Unlike at high $S/N$, negative values of $\delta_{\rm SCF}$ appear in the bottom panels. This happens because noise may make a mixed population fit slightly better than the one obtained with the correct model. In any case, a $\delta_{\rm SCF} > 0$ criterion does identify the most biased solutions.

The point to take from this experiment is that  $\delta_{\rm SCF}$ is a useful problem detector for \starlight\ runs.

\subsection{Testing previous runs}
\label{sec:TestYourRun}

As explained in Section \ref{sec:TheReason}, the biases reported by GYCMLL derive from an $A_V < 1$ initialization condition which should not affect real galaxy analysis with \starlight, but leads to very slow progress of the Markov chains in extreme situations like when $A_V$ is high and all the light is concentrated in a single population. The likelihood of real galaxy spectra triggering this anomalous behaviour of the code is small, but one would still like to have a way of testing this, and the $\delta_{\rm SCF}$ index suits this need.

To illustrate its use, we have applied this test to our $\sim 10$ yr old \starlight\ fits of SDSS DR7 spectra, used in several publications by our group and others. The information on the SCF was gathered from the \starlight's output. 
Out of 614667 galaxies in the SDSS main galaxy sample  with $S/N \ge 10$ just 798 ($\sim 0.1\%$) have $\delta_{\rm SCF} > 0$. Inspection of these cases shows that some are just bad input, but most are early-type galaxies where the mixed fit already puts most of the light in the best SCF population anyway. 
Similar results are obtained for the 363253 CALIFA spectra in the {\sc p}y{\sc casso} database \citep{Amorim2017}, 0.3\% of which have $\delta_{\rm SCF} > 0$. \starlight\ users can easily replicate this test with their own runs,
though judging from these results this is hardly necessary.

These basic tests reinforce our conclusion that the initialization which leads to the catastrophic results of GYCMLL has negligible effect in real life. Though the code is far from being perfect, it is evident that biases as large as they report are in no measure typical of \starlight.

Finally, at the risk of stating the obvious, one should always check the spectral fits. Not one-by-one, as this is not viable with the huge datasets currently available, but at least filtering those with suspicious output, like bad quality fit, low $S/N$, $\delta_{\rm SCF} > 0$, outliers in relations between derived properties, etc.  For instance, a one second look at figure 5 of GYCMLL tells a minimally experienced eye that there is something wrong with the fit. Indeed, that very figure prompted an inquire from the authors to the developer of \starlight, and all explanations given here were then offered in detail, including the revised code itself and a version of Fig.\ \ref{fig:GeFig5K0}. Yet, for unbeknownst reasons the authors chose to feature that misleading plot as an example of \starlight.

\subsection{Other experiments}

All results reported above were obtained with the same ``standard'' \starlight\ setup used in the main part of GYCMLL's paper, but they also present experiments with a ``slow mode'' configuration for the Markov chains in an  appendix. We have replicated those experiments using the revised version of the code. Results are shown in appendix \ref{app:AppendixA}. While they obtain that these fits yields much better results than those with the ``default'' configuration, we find no significant differences.

GYCMLL further explore cases where the input spectrum is not a single SSP, but a linear combination of two SSPs of different ages. This more interesting test was also replicated, and results are shown in appendix \ref{app:AppendixB}. While they find that biases in the derived properties increase to unacceptable levels as   $E(B-V)_{\rm in}$ increases, we obtain much smaller biases and no dependence on $E(B-V)_{\rm in}$.

In both cases the differences between what is reported here and there stem from the otherwise unimportant  $A_V < 1$ initialization limit of which GYCMLL were informed.

\subsection{A note on extinction}

In GYCMLL one reads that one of their motivations to study what they call the ``algorithm bias'' is that   \cite{CidFernandes2005} ``found that the dust extinction had a clear difference with that of the MPA/JHU group'' (sic). 

That is a shocking misunderstanding of section 4.1 and figure 8 in \cite{CidFernandes2005}, where the $A_z$ values derived with \starlight\ were compared to those in the value added catalogues by the MPA/JHU group \citep{Kauffmann2003a,Brinchmann2004} for SDSS galaxies. What is demonstrated and explicitly said there is exactly the opposite: ``We thus conclude that there are no substantial differences between the MPA/JHU and our estimates of the stellar extinction other than those implied by differences in the reddening laws adopted in the two studies.'' Moreover, both  \cite{CidFernandes2005} and \cite{Asari2007} report convincing empirical evidence that the stellar extinction estimated by \starlight\ is reasonable.

Still regarding extinction, and given that the biases reported by GYCMLL occur for their simulations with large  $E(B-V)_{\rm in}$, it is fit to recall that most full spectral fitting codes (\starlight\ included) treat the effects of dust in a blatantly over-simplistic way. Users dealing with dusty sources should seriously ponder how this might affect their analysis. The attenuation model assumed might have a far more relevant role than the spectral fitting algorithm.

\section{Discussion}
\label{sec:Discussion}

%------------------------------- Figure -------------------------------%
\begin{figure*}
  \includegraphics[width=\textwidth, clip=true, trim = 0 212 0 0]{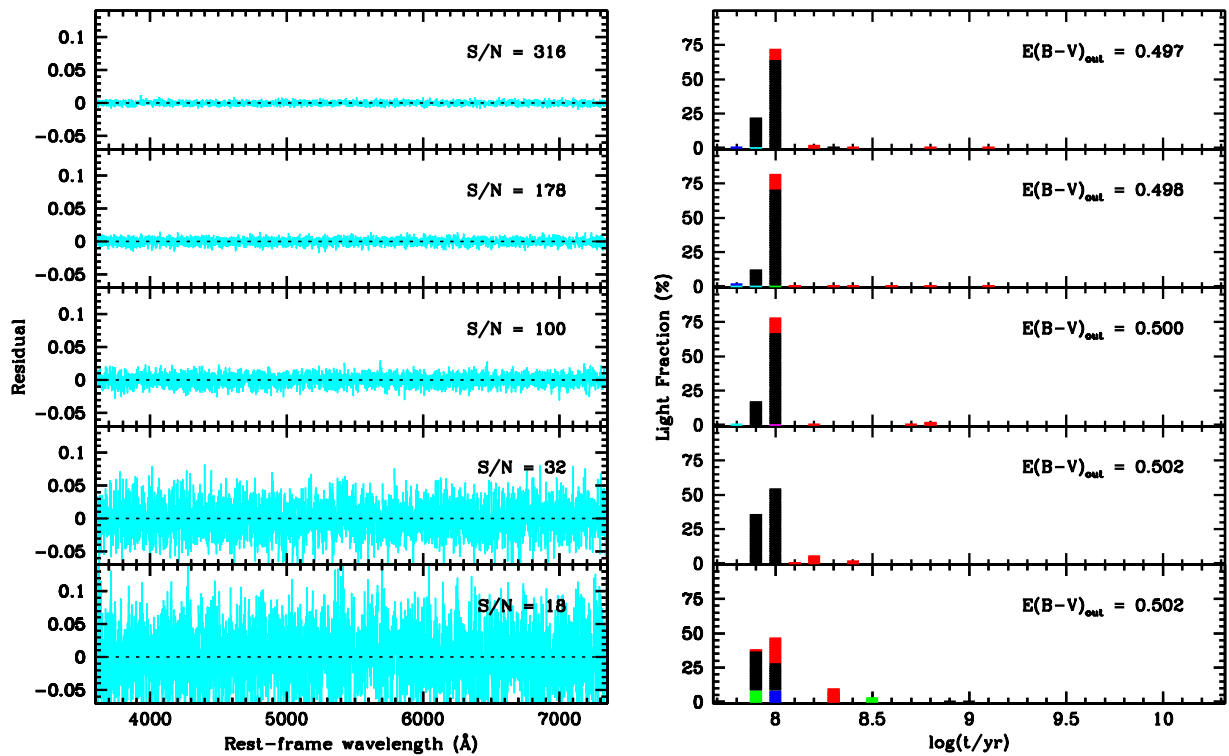}
  \caption{Output minus input residual spectra (left) and the corresponding light fractions age-distribution (right) for the same input model as in Fig.\ \ref{fig:GeFig5K0} (a 100 Myr solar SSP with $E(B-V) = 0.5$) for ``slow mode'' fits (see text). Each row is for a different level of noise.
}
\label{fig:GeFig5K2}
\end{figure*}
%------------------------------- Figure -------------------------------%

Despite the rebuttal  of the GYCMLL results presented in this communication, it is fair to recognize that their experiments lead to the identification of a previously unnoticed aspect of  the code's behaviour. As shown above, this should have no practical effect in the analysis of real galaxy data. A ``quick-and-dirty'' method to test this by comparing the best mixed model fit with the best SCF (given in the output, though seldom used) was presented and shown to corroborate this conclusion.

The code has been previously subjected to simulations qualitatively like those of GYCMLL, where the output is compared to a known input (e.g, \citealt{CidFernandes2005, CidFernandes2014}), but no serious biases were found. Recently \cite{Magris2015} compared the performance of four inverse population synthesis codes, finding that they all lead to similar accuracies in derived properties. Regarding extinction, for instance, their tests with \starlight\ yield a bias in $A_V$ of just $-0.03$ mag and a precision of $\pm 0.06$. As in our own previous simulations, \cite{Magris2015} use as input spectra computed with star formation histories and dust content designed to mimic real galaxies.

The GYCMLL simulations, on the other hand, are not based on galaxies, but reddened SSPs. This in itself already makes things difficult for a Markov chain based algorithm, as it places the right solution in an extreme corner of a vast parameter space, which may take long to reach, particularly at high $S/N$. What really triggers the anomalous behaviour of the chains, however, are the large values of extinction used in some of their tests. It is this combination of unfavorable and unrealistic conditions that sometimes turns the otherwise irrelevant $A_V < 1$ initialization limit  into a critical bottleneck, hindering the performance of the algorithm to the point that in order to obtain adequate results the setup chain parameters require adjustments, brutally slowing the code. The previous tests of the code were made for far more realistic conditions, which explains why the potential effects of the apparently (and in most cases truly) harmless  $A_V < 1$ initialization constraint were not realized before.

The only difference between the results shown in Figs.\ \ref{fig:GeFig2} and \ref{fig:GeFig5K0} with respect to their twins in GYCMLL is that they were run without the $A_V < 1$  condition. This banal modification heals all the problematic  behaviour incorrectly used to characterize \starlight's performance. 

Though that seems hardly necessary given the already excellent results shown in these plots, even better fits can be obtained by adjusting the configuration parameters. Fig.\ \ref{fig:GeFig5K2} exemplifies this by showing a series of ``slow mode'' fits obtained for the poster child case explored by GYCMLL, namely an input SSP of 100 Myr, $[M/H] = 0$ and $E(B-V) = 0.5$ observed at different $S/N$. This plot looks the same whether using the original or the revised version of the code, confirming that with appropriate parameters the Markov chains can beat the  $A_V < 1$ initialization even in this extreme case. The only difference is in the computing times, which are an order of magnitude faster with the revised code.

\section{Concluding remarks}

A common nightmare among those who offer their codes to public use is that they will be misused or unfairly characterized. It is, after all, the user's responsibility to understand the workings and the limitations of the code and to use it reasonably.  Unfortunately none of these unspoken rules-of-the-trade were followed by GYCMLL. Despite being explicitly informed on the practical irrelevance of the unexpected effect  they came across, they proceeded to publish a paper which seeds unfounded doubts not only on \starlight\ users but also on the reliability of all results obtained with the code. This is what prompted this (otherwise unnecessary) rebuttal article.

None of what was presented here implies that \starlight\ is better or worse than any other full spectra fitting code. \starlight\ does have known limitations, as consistently pointed out in our own publications. Its built-in tendency to break the solution among many components, for instance, needs to be dealt with carefully to avoid over-interpretations, as is its option to work on an observationally-oriented light fractions space instead of the more theory-guided alternative to fit for mass fractions. Smoothing the output age and metallicity arrays or  censoring components with small light fractions are  examples of post-processing steps to handle this issue. Other codes tackle this problem not a posteriori, but imposing smooth solutions to begin with, as in the pPXF-based study by \cite{McDermid2015}, where regularization in the mass fractions space is imposed. 
MOPED  \citep{Heavens2000,Panter2003}, 
STECKMAP \citep{Ocvirk2006a},
sedfit \citep{Walcher2006},
VESPA \citep{Tojeiro2007},
ULySS \citep{Koleva2009},
TGASPEX and DynBaS3D \citep{Magris2015}, 
BEAGLE \citep{Chevallard2016},
and FIREFLY \citep{Wilkinson2017}
are examples of full spectral fitting codes which offer alternative takes on this and numerous other issues involved in the non-trivial task of turning spectroscopic data on galaxies onto information on their stellar populations and dust content.

The inherent astrophysical and mathematical complexity of the subject, coupled to the numerous technicalities and philosophical issues involved in any one full spectral fitting software make the comparison of different codes a challenging task. We concur with GYCMLL in that such comparisons should be stimulated, but, as demonstrated here, this requires a level of  care that is just not met by their own study.

We close with a sobering thought. The whole area of full spectral fitting is nowhere near the standard of code-comparison achieved after decades of work by the photoionization community \citep{Ferland2016}, for instance. In fact, it is not at all clear that such a standard can even be met given the built-in differences in extant approaches to the problem. 
On the other hand, the field does have perspectives other than algorithm testing. A particularly promising line of work is to extend the wavelength range.
To mention a recent example, \cite{LopezFernandez2016} perform a combined analysis of spectra and photometry, simultaneously fitting optical spectra from the CALIFA survey and UV photometry from GALEX. The extended baseline allows one to harness valuable information from the $\lambda$-dependent behaviour of different stellar populations, ultimately leading to better constrained parameters. Studies seeking to improve upon the over-simplistic way dust attenuation is treated in full spectral fitting codes should also be encouraged.

\section*{Acknowledgements}

The author is in debt with all those who helped with \starlight\ over the years, from the students who patiently tested it to all collaborators whose input (specially when critical) contributed to keeping it up and running for so long. Support from CNPq is duly acknowledged.

\bibliography{references}

\appendix

\section{Slow mode runs}
\label{app:AppendixA}

%------------------------------- Figure -------------------------------%
\begin{figure*}
  \includegraphics[width=\textwidth, clip=true, trim = 0 170 0 0]{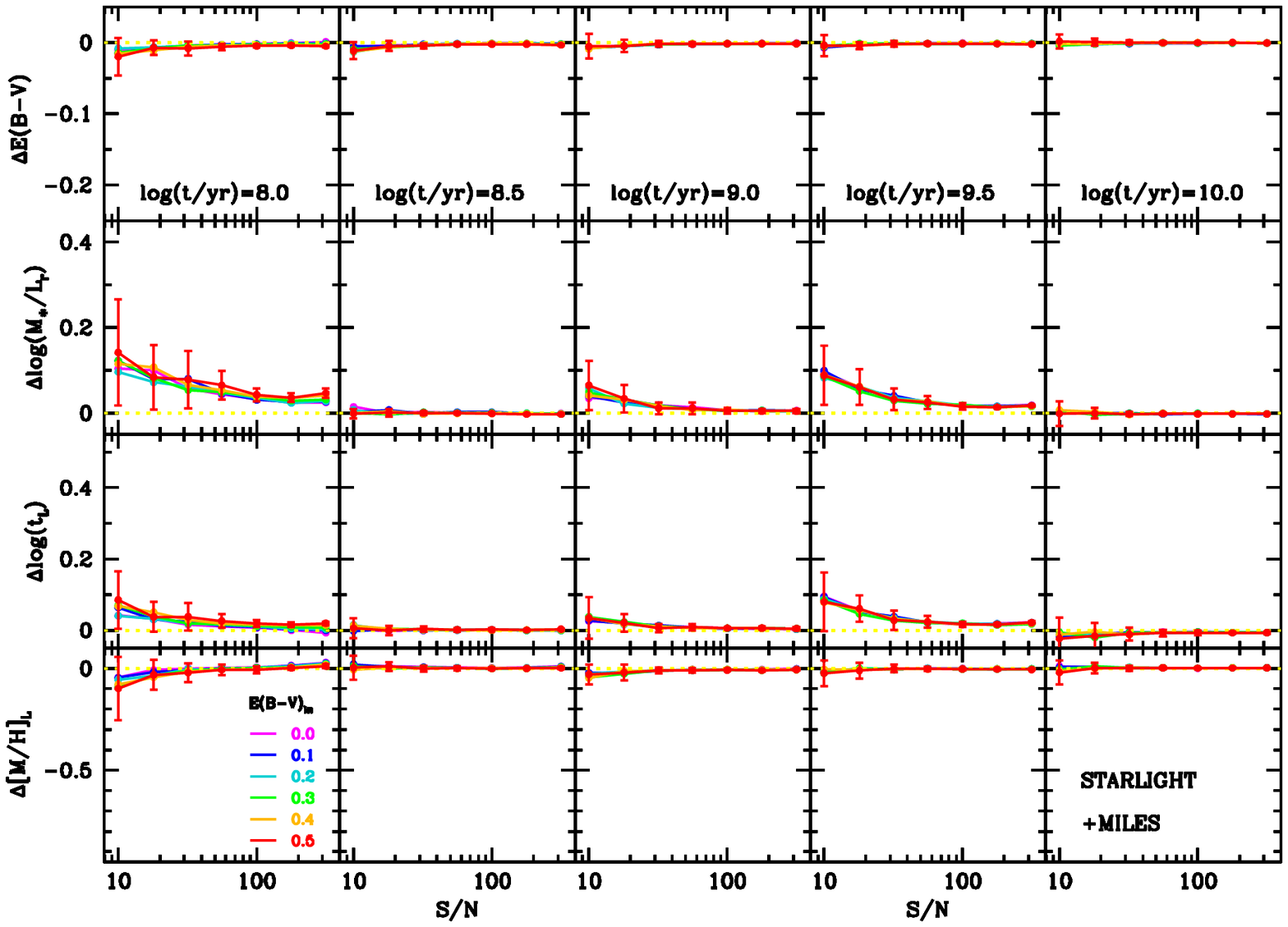}
  \caption{A revised version of figure A1 in GYCMLL. The plot is identical to Fig.\ \ref{fig:GeFig2} in the main text, except that a slow mode configuration for the Markov chains was used for the \starlight fits. Differences with respect to the results in Fig.\ \ref{fig:GeFig2} are however negligible. 
}
\label{fig:GeFig2K2}
\end{figure*}
%------------------------------- Figure -------------------------------%

GYCMLL also explore ``slow mode'' \starlight\ fits, meaning fits where the technical parameters controlling the behaviour of the Markov chains are set to allow for more chains and also to let them run for longer. They find that with this more thorough exploration of the parameter space ``the spectral fitting results show significant improvements'',  particularly for runs with $E(B-V)_{\rm in} > 0.2$.

Whilst that is in fact true with the original version of the code, by simply removing the $A_V < 1$ imposed upon initialization one finds that these improvements go from significant to negligible. This is shown in Fig.\ \ref {fig:GeFig2K2} (to be compared with figure A1 in GYCMLL), which replicates Fig.\ \ref {fig:GeFig2}, except that now the the slow mode configuration described by GYCMLL was used in the fits.

We note in passing that, despite what is said in GYCMLL, clipping\footnote{Clipping is a \starlight\ feature designed to ignore pixels which ``refuse'' to be well matched by the model, like when an emission line is not properly masked, or some bad pixel is not appropriately flagged as such.} has nothing to do with any of this.

\section{Fits for combinations of two different SSPs}
\label{app:AppendixB}

%------------------------------- Figure -------------------------------%
\begin{figure}
  \includegraphics[width=0.5\textwidth, clip=true, trim = 0 20 0 0]{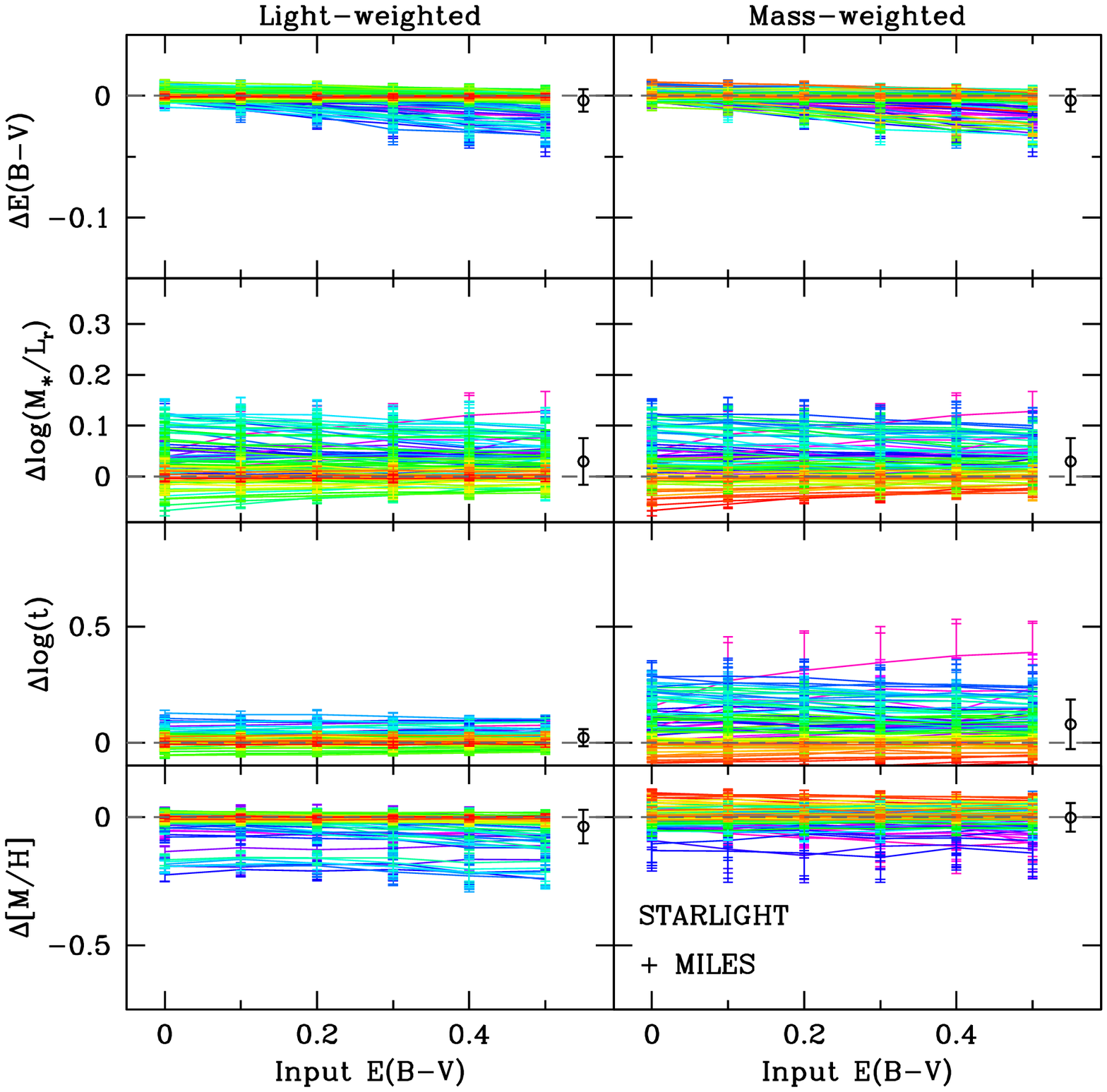}
  \caption{A revised version of figure 11 in GYCMLL, showing biases resulting from fits to a combination of two  SSPs of different ages. Each line connects the average $\Delta X = X_{\rm out} - X_{\rm in}$ values as a function of $E(B-V)_{\rm in}$, where $X$ is a given derived property. The statistics is done with 50 realizations of the noise (at $S/N = 60$). Colours code for the mean $\log (t/yr)$ of the two populations, running chromatically from 7.9 (magenta) to 10.2 (red). In the left panels the colors are coded according to the luminosity weighted mean log age, while on the right the colors map the mass weighted log age.
The open circle and error bar at the right end of the plots mark the mean and $\pm 1$ sigma range of $\Delta X$ over all 78 combinations and 50 realizations.
}
\label{fig:2cpTestsK0}
\end{figure}
%------------------------------- Figure -------------------------------%

%------------------------------- Figure -------------------------------%
\begin{figure}
  \includegraphics[width=0.5\textwidth, clip=true, trim = 0 20 0 0]{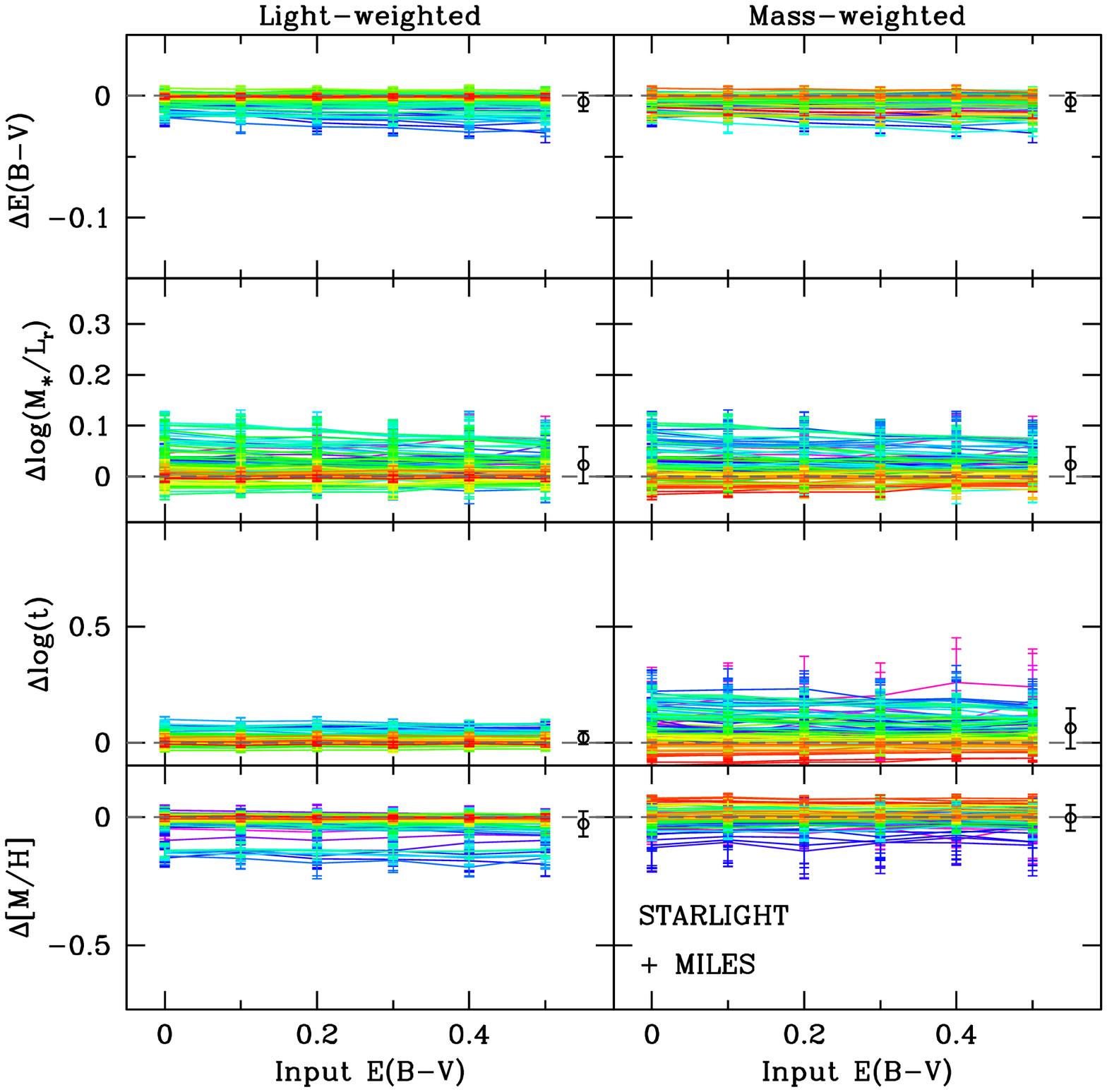}
  \caption{As Fig.\ \ref{fig:2cpTestsK0}, but for slow mode runs (analogous to figure A4 in GYCMLL). Improvements with respect to the ``standard mode'' runs in Fig.\ \ref{fig:2cpTestsK0} are only marginal.}
\label{fig:2cpTestsK2}
\end{figure}
%------------------------------- Figure -------------------------------%

GYCMLL also fitted mock spectra generated by adding two SSPs scaled to have the same flux  in the 5490--5510 \AA\ window, i.e., a fifty-fifty percent mixture in terms of light fraction. Despite its simplicity, seen in comparison with the single SSP mock spectra discussed in the main text this test goes one step further in the sense of emulating realistic galaxy spectra.
The 13 SSPs used in this experiment have solar metallicity and ages from 63 Myr to 15.8 Gyr in logarithmic steps of 0.2 dex. Spectra for each of the resulting 78 unique combinations were perturbed by noise and processed through \starlight.

Fig.\ \ref{fig:2cpTestsK0} shows our results for this same experiment. The plot replicates figure 11 in their paper, and shows biases in the derived properties as a function of the input $E(B-V)$. Each panel shows 78 lines connecting the average values of the bias obtained for 50 realizations of each of the two-SSP combinations. The color reflects the mean log age of the two SSPs, weighted either by light (left panels) or mass (right). 

Once again, and for the same reasons, our results contrast with those of GYCMLL, who identify  large biases in derived properties and a strong tendency for them to increase with increasing $E(B-V)_{\rm in}$. We find small biases and no tendency with input reddening. Only $\Delta E(B-V)$ (top panels) slightly increases with $E(B-V)_{\rm in}$ when the mean age of the pair of SSPs leans towards the smaller values allowed (lines in the magenta--blue range), but even in this case the asymptotic bias is a tiny 0.02 mag, much smaller than obtained by GYCMLL. The average and standard deviation of $\Delta E(B-V)$ over all $78 \times 50$ combinations of input spectrum and noise are $-0.004 \pm  0.009$, as show by the open circle and error bar towards the right end of the top panels in Fig.\ \ref{fig:2cpTestsK0}. For the other properties in Fig.\ \ref{fig:2cpTestsK0} we find
$\Delta \log M/L_r = 0.03 \pm 0.05$, 
$\Delta \log(t_L) = 0.02 \pm 0.04$, and
$\Delta [M/H]_L = -0.04 \pm 0.07$. 
For the mass weighted mean ages and metallicities (shown in the right panels) we obtain $\Delta \log(t_M) = 0.08 \pm 0.11$, and $\Delta [M/H]_M = -0.002 \pm 0.056$. 

Fig.\ \ref{fig:2cpTestsK2} repeats the analysis of Fig.\ \ref{fig:2cpTestsK0}, but for slow mode runs, as in figure A4 of GYCMLL. The differences with respect to Fig.\ \ref{fig:2cpTestsK0} are marginal, as can be appreciated visually comparing the two plots. In numerical terms, these longer runs yield essentially the same bias and scatter statistics:  $\Delta E(B-V) = -0.005 \pm 0.008$, $\Delta \log M/L_r = 0.02 \pm 0.04$, 
$\Delta \log(t_L) = 0.02 \pm 0.03$, 
$\Delta [M/H]_L = -0.03 \pm 0.05$, 
$\Delta \log(t_M) = 0.06 \pm 0.09$, and $\Delta [M/H]_M = -0.003 \pm 0.051$.

\end{document}